\documentclass[aps,prl,preprint]{revtex4-1}

\usepackage{amsmath}
\usepackage{amssymb}
\usepackage{graphicx}

\begin{document}

\title{Valley polarized quantum Hall effect and topological insulator phase
transitions in silicene}
\author{M. Tahir and U. Schwingenschl\"{o}gl}
\email{udo.schwingenschlogl@kaust.edu.sa,+966(0)544700080}
\affiliation{PSE Division, KAUST, Thuwal 23955-6900, Kingdom of Saudi Arabia}

\begin{abstract}
The electronic properties of silicene
are distinct from both the conventional two dimensional electron gas and the
famous graphene due to strong spin orbit interaction and the buckled
structure. Silicene has the potential to overcome limitations encountered
for graphene, in particular the zero band gap and weak spin orbit interaction.
We demonstrate a valley polarized quantum Hall effect and topological
insulator phase transitions. We use the Kubo formalism to discuss the Hall
conductivity and address the longitudinal conductivity for elastic impurity
scattering in the first Born approximation. We show that the combination of an
electric field with intrinsic spin orbit interaction leads to quantum phase
transitions at the charge neutrality point, providing a tool to experimentally
tune the topological state. Silicene constitutes a model system
for exploring the spin and valley physics not accessible in graphene due to
the small spin orbit interaction.
\end{abstract}

\maketitle

The quantum Hall effect (QHE) is one of the most striking phenomena in the
field of condensed matter physics since its discovery in the 1980s \cite{1}.
It is referred to as integer quantum Hall effect as the Hall conductivity
takes values of $2(n+1)e^{2}/h$ with an integer $n\in\mathbb{N}_0$.
The QHE in two dimensional electron gases is of particular interest.
Recently, the experimental realization of graphene, a stable
monolayer of carbon atoms \cite{2,3}, has stimulated additional interest
in two dimensional systems \cite{4,5}. Graphene exhibits quantized
conductivity values of $2(2n+1)e^{2}/h$, $n\in\mathbb{N}$.
Among the unusual transport properties, the quantum spin Hall effect is
particularly exciting as it constitutes a new phase of matter \cite{6,7}.
It requires strong spin orbit interaction (SOI) but not an external perpendicular
magnetic field. When Kane and Mele \cite{6} in a ground breaking study of
graphene had proposed a new class of insulators, the topological insulators
(TIs), great experimental and theoretical excitement was generated \cite{8,9,10,11,12,13}.
In addition to the quantum spin Hall effect, an analogous quantum valley Hall effect
\cite{14,15} arises from a broken inversion symmetry, where Dirac fermions in
different valleys flow to opposite transverse edges when an in-plane
electric field is applied in the presence of intrinsic SOI. The quantum valley
Hall effect paves the way to electric generation and detection of valley
polarization. Since the occurrence of conducting surface states in TIs is related
to the SOI \cite{12} and SOI is also a crucial criterion for the quantum
spin Hall effect, it has been proposed to search for new materials
with strong SOI for application in spintronic devices \cite{6,12}.

Silicene is a monolayer of silicon \cite{16,17} (isostructural to graphene)
with very strong SOI. It has a buckled honeycomb structure, see Fig.\ 1, where the
charge carriers behave like massless Dirac fermions \cite{18}. This breaks
the inversion symmetry and gives rise to a quantum valley Hall effect
\cite{27}. Experimental realizations of silicene sheets \cite{18,19,20} and
ribbons \cite{21,22} have been demonstrated by synthesis on metal surfaces.
It is believed that silicene opens new opportunities for electrically
tunable electronic devices \cite{23}. Although graphene possesses
extraordinary properties, its application in device fabrication is limited
by the zero band gap and the difficulty to tune the Dirac particles
electrically. Moreover, even if a band gap could be introduced by chemical
doping, it would be incompatible with existing nanoelectronics. In the desire
to overcome this limitation, the buckling in silicene offers a solution for
manipulating the particle dispersion to achieve an electrically tunable band
gap. In addition to the latter, silicene has a relatively large band gap
of 1.55 meV \cite{24} induced by intrinsic SOI, which provides a mass to the
Dirac fermions. This mass can be controlled experimentally by an external
perpendicular electric field.

In the light of the above discussion, silicene is likely to show significant
signatures of QHE as well as TI quantum phase transitions. The QHE is the
fundamental transport process under an external perpendicular magnetic field.
For intrinsic SOI and an external perpendicular electric field, we show in the
following that the quantum phase transition from a two dimensional TI to a
trivial insulator is accompanied by quenching of the QHE and onset of a
valley quantum Hall effect (VQHE), providing a tool to experimentally tune the
topological state of silicene. We use the Kubo formalism to discuss the Hall
conductivity and address the longitudinal conductivity for elastic impurity
scattering in the first Born approximation.

\section*{Results}

We model silicene by an effective Hamiltonian in the xy-plane. An external
magnetic field (0, 0, $B$) is applied perpendicular to the silicene sheet,
taking into account SOI and electric field \cite{27,25,26}. Dirac fermions in buckled
silicene obey the two-dimensional graphene-like Hamiltonian
\begin{equation}
H_{s}^{\eta}=v(\sigma_{x}\mathbf{p}_{x}-\eta\sigma_{y}\mathbf{p}_{y})-\eta
s\Delta_{SO}\sigma_{z}+\Delta_{z}\sigma_{z}. \label{1}%
\end{equation}
Here, $\eta=+/-$ denotes $K/K'$, $\Delta_{z}=lE_{z}$, where $E_{z}$ is the
uniform electric field applied perpendicular to the silicene sheet, with
$l=0.23$ \AA. In addition, ($\sigma_{x}$, $\sigma_{y}$, $\sigma_{z}$) is the
vector of Pauli matrices and $v$ denotes the Fermi velocity of the Dirac
fermions. Spin up ($\uparrow$) and down ($\downarrow$) is represented by
$s=+1$ and $-1$, respectively. Moreover, $\mathbf{p=p}+e\mathbf{A/}c$ is the two
dimensional canonical momentum with vector potential $\mathbf{A}$ and $c$ is
the speed of light. Using the Landau gauge with vector potential (0,
$Bx$, 0) and diagonalizing the Hamiltonian given in Eq.\ (1) we obtain the
eigenvalues
\begin{eqnarray}
E_{s,n,t}^{\eta } &=&t\sqrt{n\hslash ^{2}\omega ^{2}+(\Delta _{SO}-\eta
s\Delta _{z})^{2}}\quad\text{for }n>0  \label{2} \\
E_{s,0}^{\eta } &=&-(s\Delta _{SO}-\eta \Delta _{z}) \notag
\end{eqnarray}
Here, $t=+/-$ denotes the electron/hole band, $\omega =v\sqrt{2eB/\hslash }$,
and $n$ is an integer. The eigenfunctions for the $K'$ point can
be obtained by exchanging the electron and hole eigenstates in the $K$ point
solution with $\phi_{n-1}$ interchanged by $\phi_{n}$. For more details
see the supplementary information (Section I).

In the presence of a magnetic field there are two contributions to the
magnetoconductivity: the Hall and longitudinal conductivities.
The latter is the localized state contribution responsible for Shubnikov
de Haas (SdH) oscillations. The Hall conductivity is the non-diagonal
contribution. In order to calculate the electrical conductivity in the
presence of SOI, an electric field, and a perpendicular magnetic field,
we employ the general Liouville equation \cite{28}. The Hall conductivity $\sigma_{xy}$
is obtained from the non-diagonal elements of the conductivity tensor as \cite{29,30}
\begin{equation}
\sigma _{xy}=\frac{i\hslash e^{2}}{L_{x}L_{y}}\underset{\xi \neq \xi
'}{\sum }f(E_{\xi })[1-f(E_{\xi '})]\left\langle \xi
\right\vert v_{x}\left\vert \xi '\right\rangle \left\langle \xi
'\right\vert v_{y}\left\vert \xi \right\rangle \frac{1-\exp (\frac{%
E_{\xi }-E_{\xi '}}{k_{B}T})}{\left( E_{\xi }-E_{\xi ^{\prime
}}\right) ^{2}}.  \label{3}
\end{equation}%
Solving and simplifying, see the supplementary information (Section II),
yields in the limit of zero temperature
\begin{equation}
\sigma _{xy}=\frac{2e^{2}\sin ^{2}\theta _{\eta,s}}{h}\left( 2n+1+2(\frac{%
\Delta _{SO}-s\eta \Delta _{z}}{\hslash \omega })^{2}\right),  \label{5}
\end{equation}
with $\theta_{\eta,s}=\tan^{-1}\frac{\hbar\omega\sqrt{n}}{(\Delta_{SO}-\eta
s\Delta_{z})}$. This result is identical to the integer quantum Hall effect in graphene 
\cite{2,3,29,30} for $\Delta _{z}=\Delta _{SO}=0$, where the
plateaus appear at $\pm 2,\pm 6,\pm 10,...$ $\frac{e^2}{h}$, as shown in Fig.\ 2. For
$\Delta _{z}=0$ and $\Delta _{SO}\neq 0$ there is a plateau at the
charge neutrality point (CNP) in Fig.\ 2(top), which confirms that silicene is gapped
due to strong SOI. The parameters used in Figs.\ 2 to 5 are $N=1\times 10^{15}$ m$^{-2}$,
$\mu_{B}=5.788\times 10^{-5}$ eV/T, $n_{e}=5\times 10^{15}$ m$^{2}$,
$k_{0}=10^{-7}$ m$^{-1}$, $v=5\times 10^{5}$ m/s, and $\epsilon _{r}=4$. 

Furthermore, we show the Hall conductivity in Fig.\ 3(top) as a function of the
Fermi energy for fixed values of the magnetic field and temperature. We find
plateaus in the Hall conductivity at $0,\pm 1,\pm 3,\pm 5,...$ $\frac{e^2}{h}$
for both valleys in the limit of large electric field energy ($\Delta _{z}>\Delta _{SO}$).
The total Hall conductivity of the K and K$'$ valleys together for
$\Delta _{z}>\Delta _{SO}>0$ and $\Delta _{z}=\Delta _{SO}$ is presented in
Fig.\ 4(top). The extra plateaus at 0 and $\pm 1$ $\frac{e^2}{h}$ reflect
the quantum phase transition which lifts the four-fold degeneracy of the $n=0$
Landau level (LL). This is not possible in graphene. Moreover, for
$\Delta _{z}=\Delta _{SO}=4$ meV we find a single peak at the CNP with
electron hole symmetry and plateaus at $\pm 1,\pm 3,\pm 5,...$ $\frac{e^2}{h}$.
This corresponds to the semimetallic state of the system \cite{31}. There
are three major differences between silicene and graphene: First, the
electrical effects due to buckling polarize all LLs. Second,
the SOI is much stronger. Third, the spin and valley degeneracy
factor 4 is missing in the prefactor of Eq.\ (4). This demonstrates
that the surface states of the silicene TI have a strong spin texture
structure, which is distinct from the ideal Dirac fermions in graphene and
results in a different quantum Hall conductivity. We note that the $n\neq 0$
LLs are still doubly degenerate, consisting of spin up and down states from
different valleys. The factor of 2 in the prefactor of Eq.\ (4) is due to
this degeneracy.

We argue that silicene undergoes a quantum phase transition from a
trivial insulator, in which the Hall conductivity at the CNP is zero, to
a Hall insulator, in which the Hall conductivity equals
$e^{2}/h$. The transition happens when $\Delta _{z}>\Delta _{SO}>0$ and
is associated with a non-analytic contribution to the conductivity from
the $n = 0$ LL. The transition is the result of a change
in the character of the $n = 0$ LL, see Eq.\ (2), which happens in each
valley independently. For $\Delta_z=0$ one of the $n = 0$ sublevels is
electron-like and the other hole-like in both valleys. When the electric field
energy exceeds the SOI, both sublevels are electron-like for the K valley
and hole-like for the K$'$ valley. The Hall conductivity in
Eq.\ (4) evaluated for zero Fermi energy jumps from 0 ($\Delta _{z}=0$,
$\Delta_{SO}>0$) to $\pm e^{2}/h$ ($\Delta _{z}>\Delta _{SO}$) by tuning the
electric field.

To obtain the longitudinal conductivity, we assume that
the electrons are elastically scattered by randomly distributed charged
impurities, as it has been shown that charged impurities play the key role in
the transport in silicene near the Dirac point. This type of
scattering is dominant at low temperature. If there is no spin degeneracy,
the longitudinal conductivity is given by \cite{28,29,30} 
\begin{equation}
\sigma _{xx}^{\mathrm{long}}=\frac{e^{2}}{L_{x}L_{y}k_{B}T}\underset{\xi ,\xi
'}{{\displaystyle\sum }}f(E_{\xi })[1-f(E_{\xi '})]W_{\xi
\xi '}(E_{\xi },E_{\xi '})(x_{\xi }-x_{\xi '})^{2}.
\label{6}
\end{equation}%
Here, $f(E_{\xi})$ is the Fermi Dirac distribution function, with $f(E_{\xi
})=f(E_{\xi'})$ for elastic scattering, $k_{B}$ is the Boltzmann
constant, and $E_{F}$ is the chemical potential. $W_{\xi\xi'}(E_{\xi
},E_{\xi'})$ is the transmission rate between the one-electron states
$\left\vert \xi\right\rangle $ and $\left\vert \xi'\right\rangle $,
and $e$ is the charge of the electron. Conduction occurs by transitions
through spatially separated states from $x_{\xi}$\ to $x_{\xi'}$,
where $x_{\xi}=\left\langle \xi\right\vert x\left\vert \xi\right\rangle $.
The londitudinal conductivity arises as a result of migration of the cyclotron
orbit due to scattering by charged impurities. A detailed derivation is given
in the supplementary information (Section III).

In general, the oscillatory part of the longitudinal conductivity in Eq.\ (5)
is expressed as
\begin{equation}
\sigma _{xx}^{\mathrm{long}}\varpropto \underset{n,s,t,\eta }{\sum }\frac{1}{%
k_{B}T}f(E_{s,n,t}^{\eta })[1-f(E_{s,n,t}^{\eta })].  \label{7}
\end{equation}
A closer analytical examination of this result establishes an oscillation
with the SdH frequency due to the distribution function
entering the expression. For $\Delta_{z}=\Delta_{SO}=0$ we have a
symmetric electron-hole spectrum of the magnetoconductivity with a single
peak at the CNP, see Fig.\ 2(bottom). This situation is the same as for the
Dirac fermions in graphene \cite{2,3,29,30}. We obtain a gap at the CNP for
$\Delta _{z}=0$ and $\Delta_{SO}>0$ with electron-hole symmetry. For $\Delta _{z}>0$,
silicene undergoes a quantum phase transition at the CNP and a well resolved splitting
of the SdH oscillations appears. Figure 3(bottom) reveals an
oscillatory behavior with the period of the SdH oscillations for the K valley
(black line) and the K$'$ valley (red line) for $\Delta _{z}>\Delta
_{SO}>0$. This non-trivial behavior is due to the interplay of SOI and electric field,
and is highlighted in Fig.\ 3(bottom). The splitted peak is shifted to the
electron and hole region for the K and K$'$ valley, respectively, when the electric field
energy grows relative to the SOI energy. This shift of the $n=0$ LL
is consistent with the energy spectrum in Eq.\ (2). In effect, the
longitudinal conductivity at the CNP passes from a minimum to a maximum by tuning
the electric field, i.e., the system undergoes a transition from a
trivial insulator to a Hall insulator.
The behaviour can be characterized as a quantum phase transition. For $n=0$
Eq.\ (6) gives in the limit of low temperature or high magnetic field
$\sigma_{xx}^{\mathrm{long}}\varpropto \underset{s,\eta ,0}{\sum }\frac{1}{
k_{B}T}e^{-\beta \left( s\Delta _{SO}-\eta \Delta _{z}\right) }$, which is
consistent with the eigenenergy spectrum of Eq.\ (2). This fact clearly indicates
a lifting of the four-fold degeneracy of the $n=0$ LL and a quantum
phase transition at the CNP.

The total Hall (top) and longitudinal (bottom) conductivities of both valleys together
are addressed in Fig.\ 4. For $\Delta _{z}=\Delta _{SO}$, we see a splitting of the $n=0$ LL
into three peaks, one at the CNP and two with electron-hole symmetry. The energy gap between
the spin up bands closes, while the spin down bands maintain a gap.
This situation corresponds to the semimetallic state of silicene \cite{31}. To
observe the effect experimentally, the temperature broadening of the LLs must be
less than the SOI energy, which can be achieved at low temperature. Finally, we show the
total Hall and longitudinal resistivities as a function of the magnetic field
in Fig.\ 5. These results refer to the regime $\Delta _{z}>\Delta _{SO}>0$
with $\Delta _{z}=15$ meV and $\Delta _{SO}=5$ meV to cleary see the VQHE. We find
that steps between the plateaus coincide with sharp peaks of the longitudinal resistivity.
For high magnetic field, we find a significant splitting of the Hall
plateaus and the corresponding peaks in the longitudinal resistivity. In contrast,
for low magnetic field, we observe a beating pattern of the SdH oscillations
due to the energy difference between SOI and electric field for the spin up and down states. 
The electric field breaks the inversion symmetry and the surface Dirac fermions acquire
a mass due to both the SOI and electric field.

\section*{Discussion}

The following conclusions apply to both conductivities: (i) For $\Delta_z=0$ there
is a minimum at the CNP due to splitting of the $n=0$ LL in two sublevels because
of the SOI. (ii) For $\Delta_z=\Delta_{SO}$ there is a peak at the CNP with
electron-hole symmetry. This corresponds to the semimetallic state. (iii)
For $\Delta_z>\Delta_{SO}$ we have a splitting of the $n=0$ LL into four sublevels
(two peaks in the electron and two in the hole region) with a gap at the CNP. All
other LLs are also split. We call this a quantum phase transition due to
the shifting of electron and hole peaks in the K and K$'$ valley, respectively.
The four-fold degeneracy of the $n=0$ LL is fully resolved and all other LLs
split into two two-fold degenerate sublevels as the spin up states of one valley
coincide with the spin down states of the other valley. This property distinguishes
silicene from graphene \cite{2,3,29,30}.

Our analytical calculations for the Hall and
longitudinal conductivities in silicene include an electric field to
model electrical tuning in nanoelectronic applications. We have shown that the Hall
conductivity is an integer multiple of $e^{2}/h$ with plateaus at $0,\pm 1,\pm 3,\pm 5,...$
$e^{2}/h$ for $\Delta_z>\Delta _{SO}$, which is a clear signature of the VQHE. The
Hall conductivity jumps from 0 to 1 at the critical electric field of the level
crossing. This reflects the transition from a trivial insulator to a Hall insulator
at the CNP. The derived results also apply to isostructural germanene, for which the SOI
is even stronger ($\Delta _{SO}=43$ meV with $l=0.33$ \AA). Experimentally, the best
way to see the QHE and VQHE is to measure the Hall conductivity as a function of the gate
voltage which tunes the chemical potential. In silicene and germanene the
temperature will not affect the Hall plateau as the SOI and electric field are
strong. We have shown that the energy splitting due to $\Delta_z$ and $\Delta_{SO}$
leads to unconventional plateaus in the QHE and VQHE with a quantum
phase transition at the CNP. Thus, the QHE, the VQHE, and the
topological insulating states in silicene and germanene can be observed and
tuned experimentally at finite temperature. The fact that the splitting due to
the SOI can be controlled by an external electric field (gate voltage) is
of great significance for electrically tunable spintronic devices.
Our predictions represent a milestone for spin and valley electronics as
silicene enables such devices, compatible with the existing technology.

\section*{Acknowledgements}
We acknowledge fruitful discussions with M.\ Ezawa.

\section*{Author contributions}
MT performed the calculations. Both authors wrote the manuscript.

\section*{Additional Information}
\subsection*{Competing Financial Interests}
The authors declare no competing financial interests.

\begin{figure}[b]
\includegraphics[width=0.48\textwidth,clip]{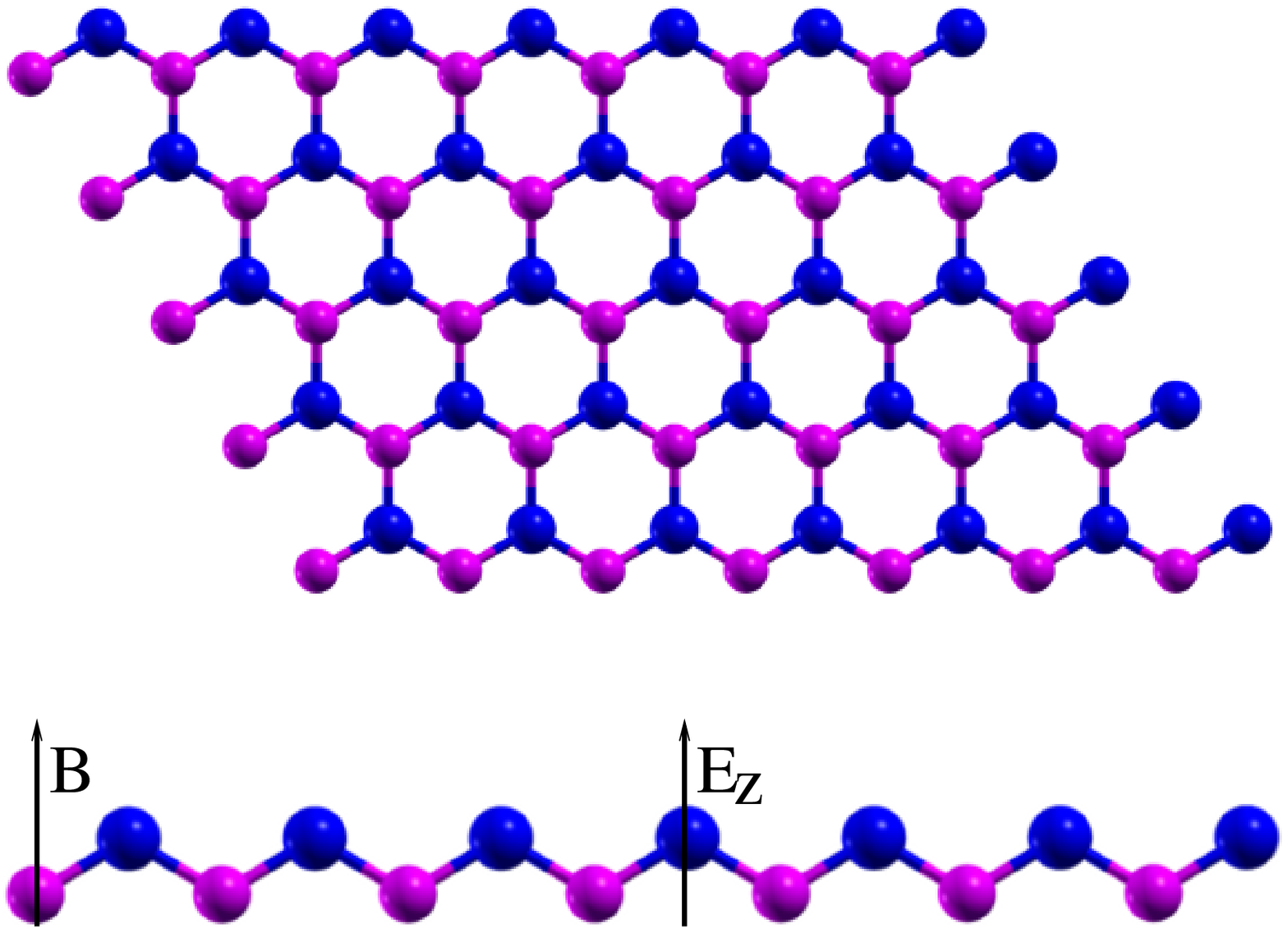}
\caption{Honeycomb lattice structure of silicene. Due to the large ionic radius
of silicon the lattice is buckled. The A and B sublattices are shifted by a
distance of $2l$ perpendicular to the silicene sheet to generate a staggered 
potential in the perpendicular electric field ($E_z$). We apply a
magnetic field ($B$) perpendicular to the silicene sheet in order to study the valley
polarized quantum Hall effect.}
\end{figure}

\begin{figure}[t]
\includegraphics[width=0.48\textwidth,clip]{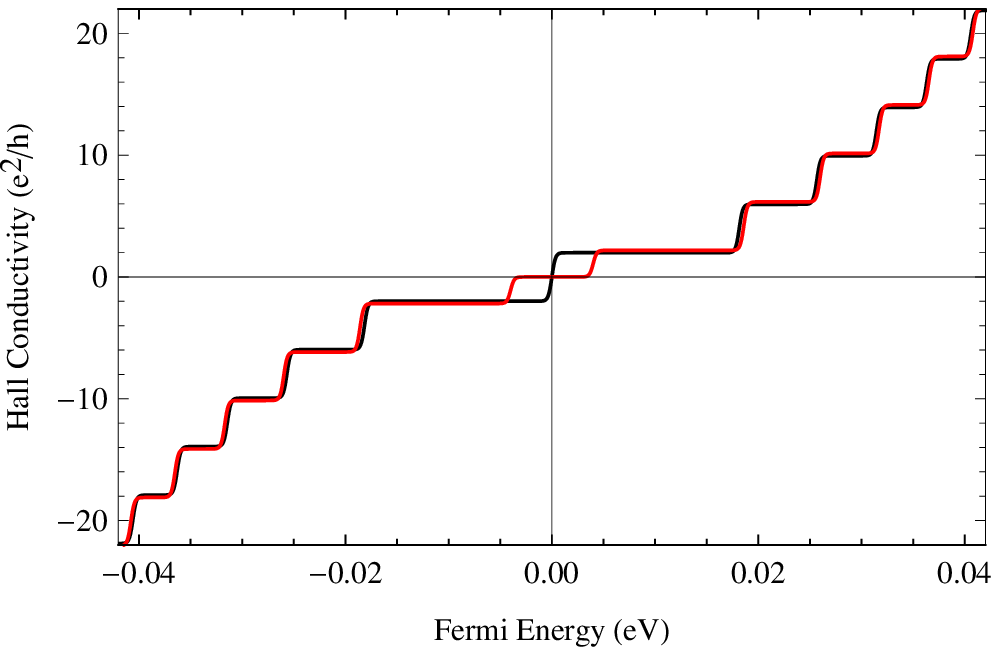}\\[0.5cm]
\includegraphics[width=0.48\textwidth,clip]{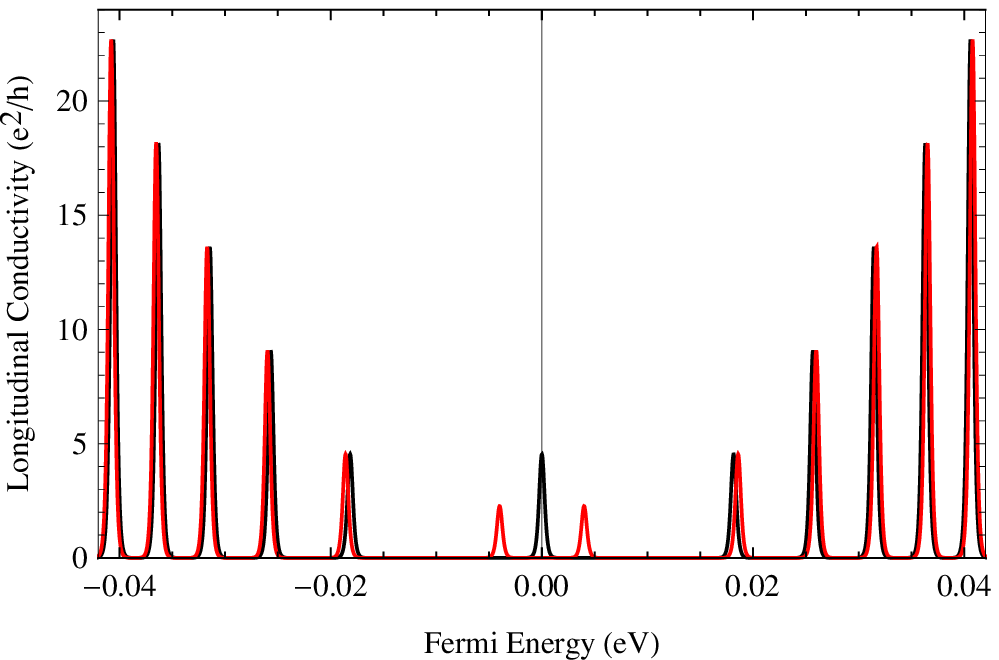}
\caption{Hall (top) and longitudinal (bottom) conductivities as functions of the Fermi energy
for 0 meV (black line) and 4 meV (red line) SOI energy.
$T=2$ K, $\Delta_z=0$ meV, and $B=1$ T. This situation represents a trivial insulator.}
\end{figure}

\begin{figure}[t]
\includegraphics[width=0.48\textwidth,clip]{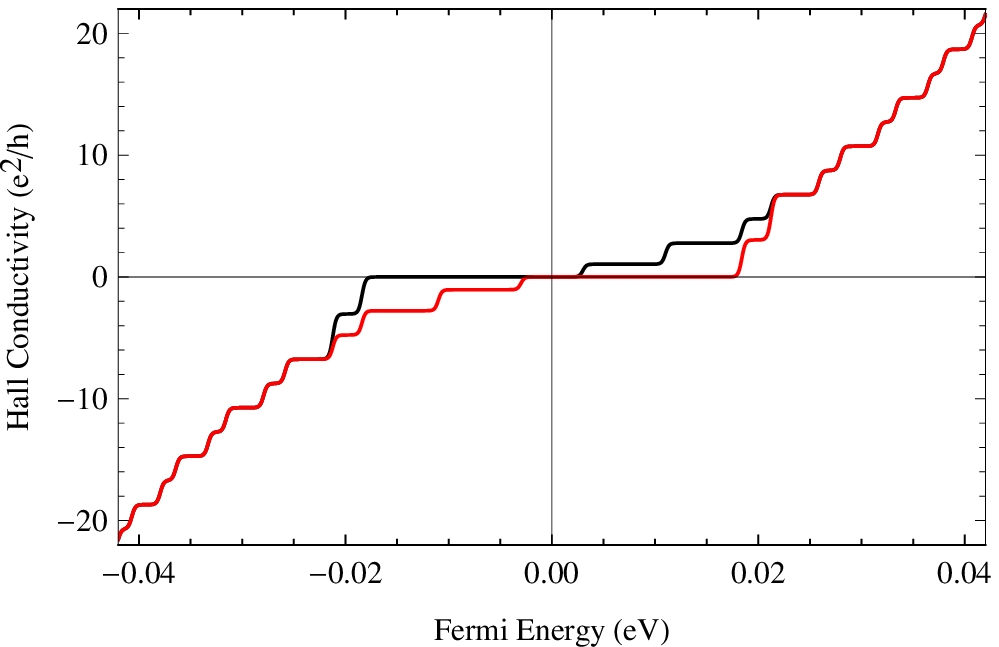}\\[0.5cm]
\includegraphics[width=0.48\textwidth,clip]{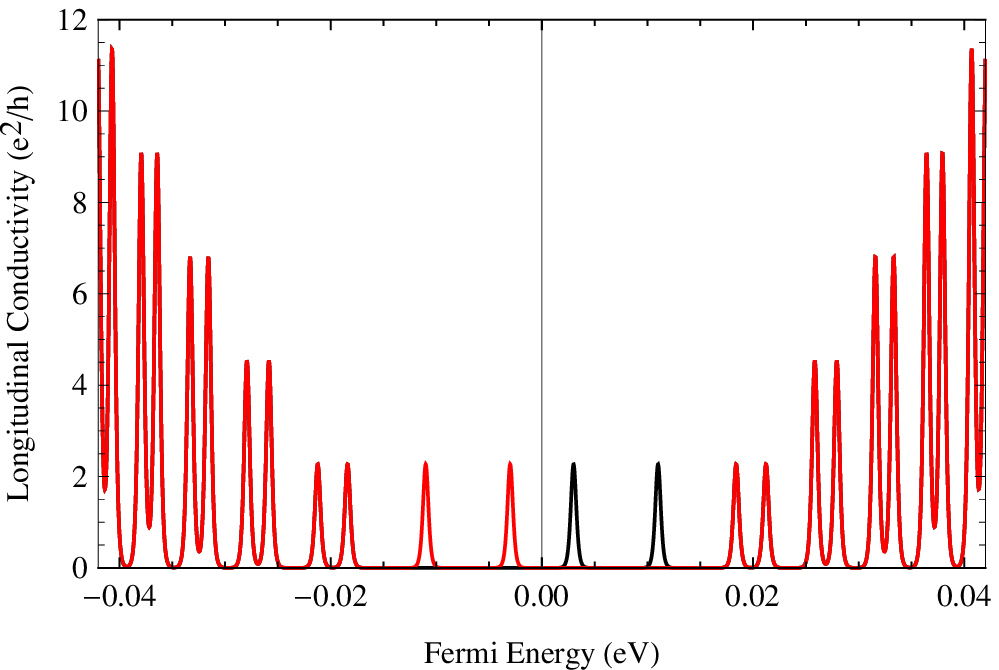}
\caption{Hall (top) and longitudinal (bottom) conductivities as functions of the Fermi energy
for 7 meV electric field energy. $T=2$ K, $\Delta_{SO}=4$ meV,
and $B=1$ T. Black lines correspond to the K valley and red lines to the K$'$ valley.
This situation represents a Hall insulator.}
\end{figure}

\begin{figure}[t]
\includegraphics[width=0.48\textwidth,clip]{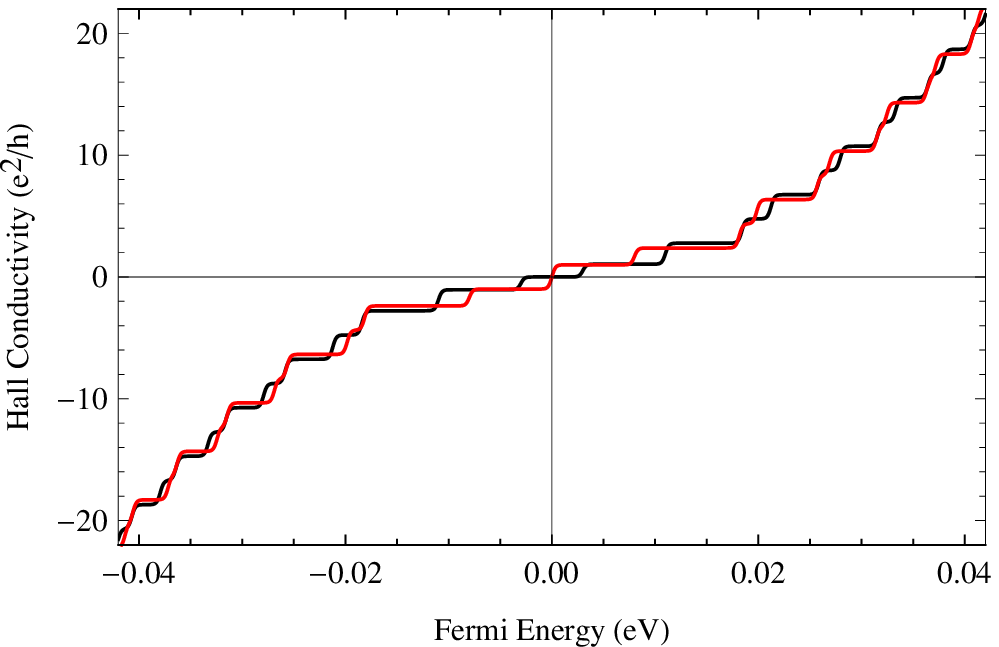}\\[0.5cm]
\includegraphics[width=0.48\textwidth,clip]{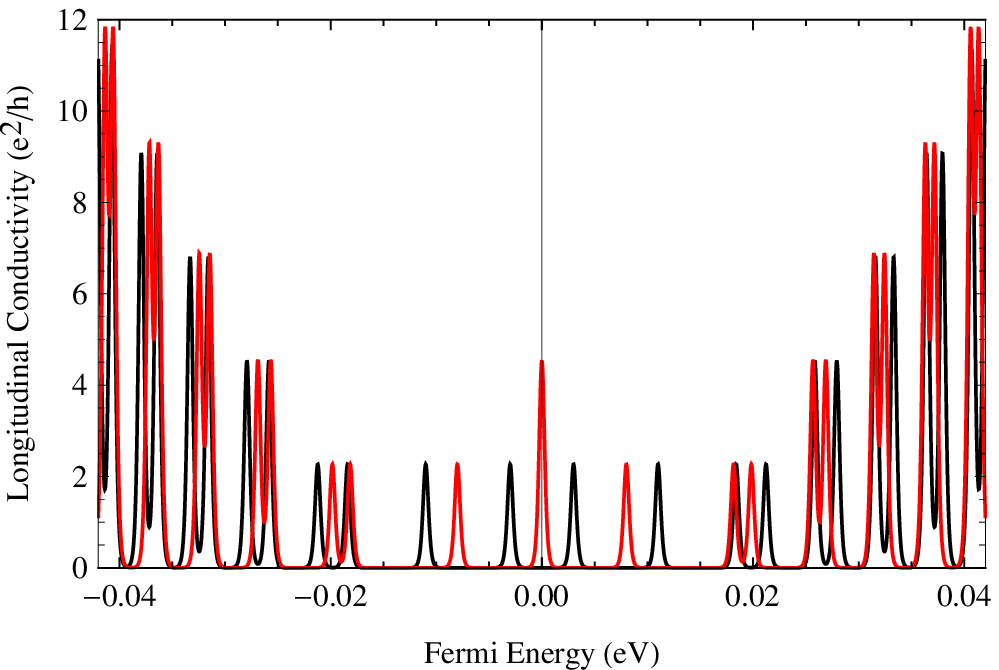}
\caption{Total (K+K$'$) Hall (top) and longitudinal (bottom) conductivities
as functions of the Fermi energy for 7 meV (black line) and 4 meV (red line)
electric field energy. $T=2$ K, $\Delta_{SO}=4$ meV, and $B=1$ T. The red
line ($\Delta_z=\Delta_{SO}=4$ meV) represents the semimetallic state.}
\end{figure}

\begin{figure}[t]
\includegraphics[width=0.48\textwidth,clip]{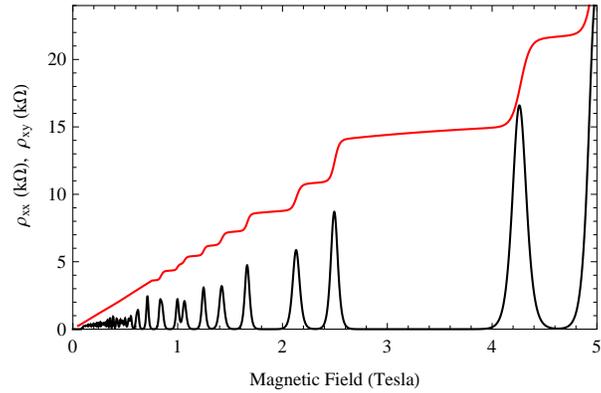}
\caption{Total (K+K$'$) Hall (red lines) and longitudinal (black lines) resistivities as functions of the
perpendicular magnetic field for $T=2$ K, $\Delta_z=15$ meV, and $\Delta_{SO}=4$ meV.
A well resolved splitting of the VQHE in the regime $\Delta_{z}>\Delta_{SO}>0$ is evident.}
\end{figure}

\end{document}